\documentclass[final,authoryear,1p,times]{elsarticle}
\usepackage{amssymb,amsmath,graphicx,epsfig,wrapfig}

\newcounter{bla}

\journal{Computer Physics Communications}

\def\hs{\hspace{-0.25cm}}
\def\hss{\hspace{-0.10cm}}

\begin{document}

\begin{frontmatter}
\title{Numerical solution of $Q^2$ evolution equations \\
       for fragmentation functions}
\author{M. Hirai}
   \address{Department of Physics, Faculty of Science and Technology,
            Tokyo University of Science \\ 
            2641, Yamazaki, Noda, Chiba, 278-8510, Japan}
\author{S. Kumano}
   \address{KEK Theory Center,
            Institute of Particle and Nuclear Studies \\
            High Energy Accelerator Research Organization (KEK) \\
            and Department of Particle and Nuclear Studies,
            Graduate University for Advanced Studies \\
            1-1, Ooho, Tsukuba, Ibaraki, 305-0801, Japan}

\begin{abstract}
Semi-inclusive hadron-production processes are becoming important
in high-energy hadron reactions. They are used for investigating
properties of quark-hadron matters in heavy-ion collisions,
for finding the origin of nucleon spin in polarized lepton-nucleon
and nucleon-nucleon reactions, and possibly for finding exotic
hadrons. In describing the hadron-production cross sections
in high-energy reactions, fragmentation functions are essential
quantities. A fragmentation function indicates  
the probability of producing a hadron from a parton
in the leading order of the running coupling constant $\alpha_s$.
Its $Q^2$ dependence is described by the standard 
DGLAP (Dokshitzer-Gribov-Lipatov-Altarelli-Parisi) evolution 
equations, which are often used in theoretical and experimental
analyses of the fragmentation functions and in calculating
semi-inclusive cross sections. The DGLAP equations are complicated
integro-differential equations, which cannot be solved in 
an analytical method. In this work, a simple method is employed
for solving the evolution equations by using Gauss-Legendre 
quadrature for evaluating integrals, and a useful code is provided
for calculating the $Q^2$ evolution of the fragmentation functions
in the leading order (LO) and next-to-leading order (NLO) of $\alpha_s$. 
The renormalization scheme is  $\overline{MS}$ in the NLO evolution.
Our evolution code is explained for using it in one's studies on
the fragmentation functions.
\end{abstract}


\end{frontmatter}

\noindent
{\bf PROGRAM SUMMARY}
 
\vspace{0.3cm}
  
\noindent
{\em Program Title: ffevol1.0}                                \\
{\em Journal Reference:}                                      \\
{\em Catalogue identifier:}                                   \\
{\em Licensing provisions: none}                              \\
{\em Programming language: Fortran77}                         \\
{\em Computer: HP DL360G5-DC-X5160}                           \\
{\em Operating system: Linux 2.6.9-42.ELsmp}                  \\
{\em RAM:} 130 M bytes                                        
{\em Classification: 11.5 Quantum Chromodynamics, Lattice Gauge Theory}  \\
{\em Nature of problem:}\\
This program solves timelike DGLAP $Q^2$ evolution equations
with or without next-to-leading-order $\alpha_s$ effects
for fragmentation functions. The evolved functions can be
calculated for 
$D_g^h$, 
$D_u^h$, $D_{\bar u}^h$, $D_d^h$, $D_{\bar d}^h$, 
$D_s^h$, $D_{\bar s}^h$, $D_c^h$, $D_{\bar c}^h$, 
$D_b^h$, and $D_{\bar b}^h$ of a hadron $h$.
   \\
{\em Solution method:}\\
The DGLAP integrodifferential equations are solved
by the Euler's method for the differentiation of $\ln Q^2$ 
and the Gauss-Legendre method for the $x$ integral
as explained in section \ref{method}.
   \\
{\em Restrictions:}\\
This program is used for calculating Q$^2$ evolution of 
fragmentation functions in the leading order or 
in the next-to-leading order of $\alpha_s$.
$Q^2$ evolution equations are the timelike DGLAP equations.
The double precision arithmetic is used.
The renormalization scheme is the modified minimal subtraction
scheme ($\overline{MS}$).
A user provides initial fragmentation functions as the subroutines
FF\_INI and HQFF in the end of the distributed code FF\_DGLAP.f.
In FF\_DGLAP.f, the subroutines are give by taking the HKNS07 
\cite{hkns07} functions as an example of the initial functions.
Then, the user inputs kinematical parameters in the file setup.ini
as explained in section \ref{input}.
   \\
{\em Running time:}\\
A few seconds on HP DL360G5-DC-X5160.
   \\


\section{Introduction}
\label{intro}

In the recent years, semi-inclusive hadron-production processes
are becoming more and more important for studying internal structure 
of hadrons and heavy-ion reactions. There are three ingredients 
for calculating their cross sections in high-energy reactions
with large transverse momenta for produced hadrons. 
The first part is on parton distribution functions (PDFs) of initial
hadrons, the second is on partonic cross sections, and the third
is on fragmentation functions (FFs) for describing production of hadrons 
\cite{ffs_before_2006, hkns07, hkos08, ffs-recent, ffs-summary}.
The perturbative aspect of quantum chromodynamics (QCD) has
been established for many processes, so that the elementary partonic
cross sections of the second part can be accurately calculated 
in high-energy processes. The PDFs of the first part have been 
extensively investigated mainly by inclusive deep inelastic scattering. 
Except for extreme kinematical conditions, the unpolarized PDFs are 
generally well determined. For example, one may look at 
Ref. \cite{recent-unpol-pdfs} for the situation of unpolarized PDFs,
Refs. \cite{aac,other-pol-pdfs} for polarized PDFs, and
Refs. \cite{our-npdfs,other-recent-npdfs} for nuclear PDFs.
Because the PDF and partonic cross section parts are relatively
well known, the only issue is the accuracy of the FFs of the third
part for calculating precise semi-inclusive cross sections.

The first estimate for uncertainties of the FFs was done in
Ref. \cite{hkns07}, and its results indicated that they have large
uncertainties especially for so called disfavored fragmentation
functions. 
This fact could add ambiguities to calculated cross sections
of high-energy hadron productions such as 
$\vec p + \vec p \rightarrow \pi+X$ and $A + A' \rightarrow h+X$
at RHIC and LHC. Here, where $\vec p$ and $\pi$ 
indicate a polarized proton and a pion, respectively, 
$X$ indicates a sum over all other hadrons created in the reaction, 
$A$ and $A'$ are nuclei, and $h$ is a produced hadron.
Furthermore, the FFs could be also used for other studies
in searching for exotic hadrons by noting characteristic
differences in favored and disfavored FFs as pointed out
in Ref. \cite{hkos08}. 

These topics suggest that the FFs should be one of most
important quantities in describing high-energy hadron reactions. 
There are two important variables in the FFs. One is the energy 
fraction $x$ (or often denoted as $z$) for a produced hadron 
from a parton and the other is the hard scale $Q^2$.
Definitions of these quantities are given in Sec. \ref{ffs}.
The $x$-dependent functions are determined mainly from
experimental measurements of electron-positron annihilation
processes $e^+ + e^- \rightarrow h+X$ by a global analysis
\cite{ffs_before_2006, hkns07, hkos08, ffs-recent}.
On the other hand, $Q^2$ dependence
can be calculated in perturbative QCD. The standard equations
for describing $Q^2$ variations are the timelike DGLAP
evolution equations \cite{dglap}.

The evolution equations are complicated integro-differential
equations, which cannot be solved by an analytical method,
especially if higher-order corrections are included in the equations.
They are solved by numerical methods. A popular method is
to use the Mellin transformation \cite{mellin}, in which
$Q^2$ evolution of moments for the FFs is analytically
calculated and resulting moments are then transformed
into $x$-dependent functions by the inverse Mellin transformation.
One of other numerical methods is to solve the $x$-integral
part by dividing the $x$ axis into small steps for calculating
integrals \cite{bfevol,trans-evol}, which is so called brute-force
or Euler method. Of course, the integral could be calculated
by a better method such as the Simpson method or Gauss-Legendre
quadrature. Another approach is to expand the FFs and
DGLAP splitting functions in terms of orthogonal polynomials
such as the Laguerre polynomials \cite{lag}.
Advantages and disadvantages of these numerical methods
are explained in Ref.\cite{kn04}.
There are also recent studies on the numerical solution
\cite{recent}.

Although the FFs are important, it is unfortunate that no useful
code is available in public for calculating $Q^2$ variations of
the FFs. For example, the FFs are calculated in a theoretical 
model \cite{bentz} at a typical hadron scale of small $Q^2$.
In order to compare with the FFs obtained by global analyses
or with experimental data, one needs to calculate the $Q^2$
evolution. However, one should make one's own code or rely on
a private communication for obtaining a code 
since no public code is available,
although the $Q^2$ evolution is often used
also in theoretical and experimental analyses. In this work, 
we explain our method for solving the DGLAP evolution and 
a useful code is supplied for public use.

This paper consists of the following.
The fragmentation functions and their kinematical variables
are introduced in Sec. \ref{ffs}, and evolution equations
are explained in Sec. \ref{q2evol}.
Our numerical method is described in Sec. \ref{method}
for solving the DGLAP $Q^2$ evolution equations, and
a developed evolution code is explained in Sec. \ref{code}.
Numerical results are shown in Sec. \ref{results} by running
the evolution code, and our studies are summarized 
in Sec. \ref{summary}.

\section{Fragmentation functions}
\label{ffs}

Fragmentation functions are given in the electron-positron annihilation
process $e^+ +e^- \rightarrow h+X$, where $h$ indicates a specific
hadron. The process is described first by a $q\bar q$ creation by 
$e^+ e^- \rightarrow q\bar q$ and a subsequent fragmentation,
namely a hadron-$h$ creation from the primary quark or antiquark.
The fragmentation function is defined by the hadron-production
cross section of $e^+ +e^- \rightarrow h+X$ as \cite{esw-book}: 
\begin{equation}  
F^h(x,Q^2) = \frac{1}{\sigma_{tot}} 
\frac{d\sigma (e^+e^- \rightarrow hX)}{dx} ,
\end{equation}
where $\sigma_{tot}$ is the total hadronic cross section.
The variable $Q^2$ is the virtual photon or $Z^{\, 0}$ momentum squared
in $e^+e^- \rightarrow \gamma$ (or $Z^{\, 0}$) and it is expressed by
the center-of-mass energy $\sqrt{s}$ as $Q^2=s$.
The variable $x$ is the hadron energy $E_h$ scaled to 
the beam energy $\sqrt{s}/2$, and it is defined by the fraction: 
\begin{equation}   
x \equiv \frac{E_h}{\sqrt{s}/2} = \frac{2E_h}{\sqrt{Q^2}}.
\end{equation}

The fragmentation process is described by the summation of hadron 
productions from primary quarks, antiquarks, and gluons \cite{esw-book}:
\begin{equation} 
F^h(x,Q^2) = \sum_i C_i(x,\alpha_s) \otimes D_i^h (x,Q^2),
\label{eqn:fh}
\end{equation}
where $C_i(x,\alpha_s)$ is a coefficient function, and it is calculated
in perturbative QCD \cite{qqbar-cross}.
The factor $\alpha_s (Q^2)$ is the running coupling constant,
and its expression is given in Appendix A 
for the leading order (LO) and next-to-leading order (NLO).
The function $D_i^h (x,Q^2)$ is the fragmentation function from a parton 
$i$ ($=u,\ d,\ s,\ \cdot\cdot\cdot,\ g$) to a hadron $h$, and
it is the probability of producing the hadron $h$, in the LO of $\alpha_s$,
from the parton $i$ with the energy fraction $x$ and the momentum
square scale $Q^2$.
The notation $\otimes$ indicates a convolution integral defined by
\begin{equation}
f (x) \otimes g (x) = \int^{1}_{x} \frac{dy}{y}
            f(y) g\left(\frac{x}{y} \right)  .
\end{equation}
The fragmentation function is formally given by
the expression \cite{ffs-def}
\begin{equation}
D_i^h (x) = \sum\limits_X \int \frac{dy^-}{24\pi } e^{ik^+ y^-}
{\rm{Tr}} \left[ {\gamma^+  
\left\langle {0 \left| {\left. {\left. \psi _i (0,y^-  ,0_\bot) \right|h,X} 
\right\rangle \left\langle {h,X\left| {\bar \psi _i (0)} \right.} \right.} 
\right|0} 
\right\rangle } \right] ,
\label{eqn:formal-ffs}
\end{equation}
where $k$ is the parent quark momentum, the lightcone notation is defined
by $a^\pm =(a^0 \pm a^3)/\sqrt{2}$, the variable $x$ is then given by
$x=p_h^+/k^+$ with the hadron momentum $p_h$, and $\perp$ is
the direction perpendicular to the third coordinate.
A gauge link needs to be introduced in Eq. (\ref{eqn:formal-ffs})
so as to satisfy the color gauge invariance. 
It should be, however, noted that a lattice QCD calculation is not 
available for the FFs because the operator-product-expansion method 
cannot be applied due to the fact that a specific hadron $h$ should 
be observed in the final state with the momentum $p_h$.

An important sum rule of the FFs is on the energy conservation.
Since the variable $x$ is the energy fraction for the produced 
hadron, its sum weighted by the fragmentation functions, 
namely the sum of their second moments, should be one.
\begin{equation}
\sum_h M_i^h = \sum_h \int_0^1 dx \, x \, D_i^h (x,Q^2) = 1 .
\end{equation}
The fragmentation function should vanish kinematically 
at $x=1$ and it is expected to be a smooth function at small $x$,
so that it is typically parametrized in the form
\cite{ffs_before_2006, hkns07, hkos08, ffs-recent}
\begin{equation}
D_{i}^{h} (x,Q_0^2) = N_{i}^{h} x^{\alpha_{i}^{h}} (1-x)^{\beta_{i}^{h}},
\end{equation}
at fixed $Q^2$ ($\equiv Q_0^2$). 
Current experimental data are not accurate enough to find
much complicated $x$-dependent functional form.
In order to calculate the function $D_{i}^{h} (x,Q^2)$ 
at arbitrary $Q^2$, one should rely on $Q^2$ evolution equations 
and the standard ones are the DGLAP equations
in the next section. 

\section{$Q^2$ evolution equations}
\label{q2evol}

The FFs depend on two variables $x$ and $Q^2$. 
The $x$-dependence is associated with a non-perturbative aspect
of QCD, so that the only way of calculating it theoretically
is to use hadron models, because the lattice QCD estimate is
not available for the FFs. There are some hadron-model calculations
\cite{bentz} by using the expression Eq. (\ref{eqn:formal-ffs}) 
at a small hadronic $Q^2$ scale. On the other hand,
$x$-dependent functions are determined by global
analyses of experimental data mainly on $e^+ +e^- \rightarrow h+X$
\cite{ffs_before_2006, hkns07, hkos08, ffs-recent}.
In the model calculations and also in the global analyses,
the $Q^2$ dependence or so called scaling violation is
calculated in perturbative QCD.

The $Q^2$ dependence of the FFs is described by the DGLAP evolution
equations in the same way with the ones for the PDFs
with slight modifications in splitting functions.
They are generally given by \cite{dglap,esw-book}
\begin{align}
\frac{\partial}{\partial \ln Q^2} D_{q^+_i}^h (x,Q^2)
& = \frac{\alpha_s (Q^2)}{2\pi} \,
  \left [
  \sum_j P_{q_j q_i}(x,\alpha_s) \otimes D_{q^+_j}^h (x,Q^2)
     + 2 P_{g q}(x,\alpha_s) \otimes D_{g}^h (x,Q^2)
  \right ] ,
\nonumber \\
\frac{\partial}{\partial \ln Q^2} D_{g}^h (x,Q^2)
& = \frac{\alpha_s (Q^2)}{2\pi} \,
  \left [
    P_{q g}(x,\alpha_s) \otimes \sum_j D_{q^+_j}^h (x,Q^2)
+   P_{g g}(x,\alpha_s) \otimes D_{g}^h (x,Q^2)
  \right ] ,
\label{eqn:evolution-q+}
\end{align}
where $D_{q^+}^h (x,Q^2)$ denotes the fragmentation-function
combination $D_q^h (x,Q^2) +  D_{\bar q}^h (x,Q^2)$. If the sum
is taken over the flavor, it becomes the singlet function
$D_{q_s}^h (x,Q^2)=\sum_q [ D_q^h (x,Q^2) +  D_{\bar q}^h (x,Q^2) ]$,
and $N_f$ is the number of quark flavors. 
The flavor nonsinglet evolution, for example,
for $q-\bar q$ type function is described by
\begin{align}
\frac{\partial}{\partial \ln Q^2} D_{q_i^-}^h (x,Q^2) 
 = \frac{\alpha_s (Q^2)}{2\pi} \, \sum_j P_{q_j q_i}(x,\alpha_s)
       \otimes D_{q_j^-}^h (x,Q^2),
\label{eqn:evolution-ns}
\end{align}
where $D_{q^-_i}^h (x,Q^2) = D_{q_i}^h (x,Q^2) - D_{\bar q_i}^h (x,Q^2)$.
The functions $P_{qq}(x)$, $P_{gq}(x)$, $P_{qg}(x)$, and $P_{gg}(x)$
are time-like splitting functions, and $P_{ij}(x)$ describes the splitting
probability that the parton $j$ splits into $i$ with the momentum
fraction $x$. It should be noted that  $P_{gq}(x)$ and $P_{qg}(x)$
are interchanged in the splitting function matrix for the PDFs.
The LO splitting functions are the same as the space-like ones;
however, there are differences between them in the NLO and higher orders
\cite{esw-book,splitting}.
Actual expressions of the LO splitting functions are provided
in Appendix B. The NLO expressions should be found in 
Ref. \cite{esw-book} because they are rather lengthy.

The DGLAP equations in Eq. (\ref{eqn:evolution-q+}) are coupled 
integro-differential equations with complicated $x$-dependent functions 
especially if higher-order $\alpha_s$ corrections are taken into account. 
It is obvious that they cannot be solved in a simple analytical form.
The convolution integral is generally expressed by a simple 
multiplication of Mellin moments, so that the the equations are
easily solved in the Mellin-moment space. However, the inverse Mellin
transformation should be calculated by a numerical method in any case
to obtain the $x$-dependent function. Here, we solve the DGLAP
equations directly in the $x$ space by calculating the $x$ integral
in a numerical way. Advantages and disadvantages of both methods
are discussed in Ref. \cite{kn04}.

\vfill\eject

\section{Numerical method for solving $Q^2$ evolution equations}
\label{method}

The integro-differential equations of Eqs. (\ref{eqn:evolution-q+})
and (\ref{eqn:evolution-ns}) are solved in the following way.
Scaling violation ($Q^2$ dependence) of the FFs is roughly given
by $\ln Q^2$, which is defined as the variable $t$:
\begin{equation}
t = \ln Q^2 .
\end{equation}
Because the $t$ dependence is not complicated in the FFs,
we do not have to use a sophisticated method for solving
the differentiation. The following simple method is used
for solving the differentiation:
\begin{equation}
\frac{d f(t)}{dt} = \, \frac{f(t_{\ell+1})-f(t_\ell)}{\Delta t} .
\label{eqn:EULER}
\end{equation}
Here, the variable $t$ is divided into $N_t$ steps with 
a small interval $\Delta t$. This method could be
called Euler method \cite{SIMP}.
It is also possible to use the Euler method for the integration part
by dividing the $x$ region into $N_x$ steps with the interval
$\Delta x$ \cite{bfevol}. However, it is more desirable to use
a better method since the $x$ dependencies of 
the FFs and splitting functions are not simple.
Here, the Gauss-Legendre method is used for calculating
the integral over $x$:
\begin{equation}
\int_{x_0}^1 g(x) \ dx 
 \simeq \frac{1-x_0}{2} \sum_{k=1}^{N_{GL}} \, w_k g(x_k) ,
\label{eqn:gauss}
\end{equation}
where $x_k=[1+x_0 +(1-x_0) x_k^\prime]/2$ with
the zero points $x_k^\prime$ of the Gauss-Legendre polynomials
in the region $-1 \le x_k^\prime \le +1$,
$w_k$ are the weights \cite{recipes}, and
$N_{GL}$ is the number of Gauss-Legendre points.
In our previous works \cite{bfevol,trans-evol},
simpler methods are used for calculating the integral
by the Euler method and the Simpson's one.
Here, we change the method for the Gauss-Legendre one
for getting more accurate numerical results.

In the following, only the nonsinglet evolution 
in Eq. (\ref{eqn:evolution-ns}) is discussed
because an extension to the general evolution in 
Eq. (\ref{eqn:evolution-q+}) is obvious just by
writing down two coupled equations in the same way.
Substituting Eqs. (\ref{eqn:EULER}) and (\ref{eqn:gauss})
into the nonsinglet equation of Eq. (\ref{eqn:evolution-ns}),
we obtain
\begin{equation}
D_{q_i^-}^h (x_m,t_{\ell+1}) 
 = D_{q_i^-}^h (x_m,t_\ell) 
  + \Delta t \, \frac{\alpha_s (t_\ell)}{2\pi} \frac{1-x_m}{2}
    \sum_{j} \sum_{k=1}^{N_{GL}}  w_k \, \frac{1}{x_k} \,
    P_{q_j q_i} (x_k) \, D_{q_j^-}^h 
    \left( \frac{x_m}{x_k}, t_\ell \right) .
\label{eqn:step}
\end{equation}
In previous codes of $Q^2$ evolution equations in 
Refs. \cite{bfevol,trans-evol}, an option is provided to divide
$\ln x_{Bj}$, where $x_{Bj}$ is the Bjorken scaling variable, 
into equal steps instead of linear-$x_{Bj}$ steps because 
the small-$x_{Bj}$ region is often important in discussing 
deep inelastic structure functions.
However, the small-$x$ part is not as reliable as the PDF case
because experimental data do not exist at very small $x$
and because of theoretical issues on finite hadron masses
and resummation effects. The Gauss-Legendre points 
are taken by considering the linear-$x$ scale
at $x>0.1$ and by the logarithmic-$x$ scale at $x<0.1$.
If the initial function is supplied at certain $Q^2$ ($\equiv Q_0^2$), 
the evolution from $t_1 = ln Q_0^2$ to the next point 
$t_2 = t_1 + \Delta t$ is calculated by 
Eq. (\ref{eqn:step}). Repeating this step, we finally obtain
the evolved FF at $t_{N+1} = \ln Q^2$.

The most important and time-consuming part is to calculate the $x$
integrals by the Gauss-Legendre quadrature. 
For the integral from the minimum $x_0$ to 1, the splitting
functions $P_{q^-} (x_k)$ are first calculated at $N_{GL}$
points of $x_k$ and they are stored in an array. Then,
the fragmentation functions are also calculated at $x_k$ and $x_m$,
and they are stored in a two-dimensional array. These arrays are
used for calculating the Gauss-Legendre sum in Eq. (\ref{eqn:step}).

\section{How to run the $Q^2$ evolution code}
\label{code}

We made the numerical evolution code of the FFs by the method discussed
in the previous section. Its main code (FF\_DGLAP.f),
a test program (sample.f), and an example of the input file
(setup.ini) could be obtained upon email request \cite{request}.
There are three major steps for calculating the $Q^2$ evolution
of the FFs:
\vspace{-0.12cm}
\begin{itemize}
\item[1.] Initial FFs are supplied in the subroutines, 
          FF\_INI for gluon ($g$) and light-quark
           ($u$, $d$, $s$, $\bar u$, $\bar d$, $\bar s$) functions
          and HQFF for heavy-quark ($c$, $b$) functions.
\vspace{-0.12cm}
\item[2.] Input parameters for the evolution are supplied 
          in the file setup.ini. These parameters are used for
          calculating two-dimensional ($x$ and $Q^2$) grid data
          for the FFs in the ranges $x_{min} \le x \le 1$ and 
          $Q_0^2=Q^2_{min} \le Q^2 \le Q_{max}^2$.
\vspace{-0.12cm}
\item[3.] As indicated in the test code (sample.f), the evolved
          $Q^2$ value (Q2) and the value of $x$ (X) should be supplied
          for calculating the evolution. 
          The grid data created in the step 2 are used 
          for this final step calculation. Therefore, 
          output functions can be obtained
          at various $x$ and $Q^2$ points without repeating
          the $Q^2$ evolution calculations as far as they are
          within the ranges $x_{min} \le x \le 1$ and 
          $Q_0^2 \le Q^2 \le Q_{max}^2$.
\end{itemize}

\subsection{Main evolution code}
\label{main-code}

The main $Q^2$ evolution code (FF\_DGLAP.f) is rather long,
so that only the major points are explained.
First, one needs to supply the initial FFs in the subroutines 
FF\_INI and HQFF, which are located in the end
of FF\_DGLAP.f. The subroutine FF\_INI is for gluon ($g$) 
and light-quark ($u$, $d$, $s$, $\bar u$, $\bar d$, $\bar s$) 
functions, and HQFF is for heavy-quark ($c$, $b$) functions.
As an example, the HKNS07 functions \cite{hkns07} are given.
The initial scale for the gluon and light-quark functions
is $Q_0^2$, and the scales are the mass-threshold
values $m_c^2$ and $m_b^2$ for charm and bottom FFs, respectively.
These scale values are provided in setup.ini, and the initial functions
are supplied in analytical forms in our main code FF\_DGLAP.f.

Second, input parameters are read from setup.ini, which
is explained in Sec. \ref{input}. They are basic parameters:
the order of $\alpha_s$, scale parameter of QCD ($\Lambda$),
charm and bottom masses $m_c$ and $m_b$ for setting
thresholds, number of flavors at the initial scale $Q_0^2$;
kinematical parameters: initial scale $Q_0^2$, maximum $Q^2$ value
$Q^2_{max}$, and minimum $x$ ($x_{min}$) for making grid data 
of the evolved FFs;
parameters to control the numerical integrations:
Gauss-Legendre points ($N_{GL}$) and numbers of $\ln Q^2=t$
and $x$ points ($N_t$ and $N_x$).

Third, the splitting functions are calculated at the points $x_k$
for calculating the summation in Eq. (\ref{eqn:step}). The $x_k$
points are determined by the parameters $N_x$ and $N_{GL}$.
The splitting functions at these points are calculated at once
in the beginning of this code. In the same way, the initial
FFs are also calculated at the given points of $x_k$ and $x_m$
and they are stored in two-dimensional ($k$ and $m$) arrays.

Forth, the evolution step of Eq. (\ref{eqn:step}) is repeated
in $N_t$ times to obtain the evolved FFs up to $Q_{max}^2$
from $Q_0^2$ in the range $x_{min} \le x \le 1$. 
If $Q^2$ exceeds the threshold $m_c^2$ (or $m_b^2$),
the number of flavor is changed accordingly and charm (or bottom) 
function starts to participate in the evolution calculation.
During the evolution calculations, two dimensional ($x$ and $Q^2$)
grid data are stored for calculating the FFs at any point
within the ranges $x_{min}<x<1$ and $Q_0^2 < Q^2 < Q_{max}^2$
by interpolation. The $x$ and $Q^2$ values need to be specified
in running this main subroutine, and an example is proved as
a test code (sample.f).

\subsection{Input file}
\label{input}

The input parameters should be supplied in the file setup.ini
for running the main evolution routine FF\_DGLAP.f, in which
the parameter values are read. For example, the following
input values are used for evolving the HKNS07 FFs in the NLO.
Here, the symbol \# is for commenting out 
the subsequent line in setup.ini.

\begin{alignat}{3}
& \text{\# pQCD ORDER 1:LO, 2:NLO}          & \ \ \ \ \ \ &
\nonumber \\
& \text{IORDER= 2}                          & \ \ \ \ \ \ &
\nonumber \\
& \text{\# DLAM (Scale parameter in QCD) of $N_f=4$}   & \ \ \ \ \ \ &
\nonumber \\
& \text{\# e.g. 0.220 GeV (LO), 0.323 GeV (NLO) in HKNS07} & \ \ \ \ \ \ &
\nonumber \\
& \text{DLAM= 0.323}           &       &   \text{\# in HKNS07-NLO}
\nonumber \\
& \text{\# Heavy-quark mass threshold}      & \ \ \ \ \ \ &
\nonumber \\
& \text{\# HQTHRE= $m_c$, $m_b$ = 1.43, 4.3 GeV in HKNS07}
                                       & \ \ \ \ \ \ &
\nonumber \\
& \text{HQTHRE= 1.43,  4.3}    &       & 
\nonumber \\
& \text{\# Q2 range for making grid files}  & \ \ \ \ \ \ &
\nonumber \\
& \text{\# Q02 $\rightarrow$ Q2max 
            (note: not the Q2 evolution range)} & \ \ \ \ \ \ &
\nonumber \\
& \text{Q2= 1.D0, 1558.D+5}    &       &   \text{\# in HKNS07 Library}
\nonumber \\
& \text{\# minimum of $x$}                    & \ \ \ \ \ \ &
\nonumber \\
& \text{XMIN= 1.D-2}                        & \ \ \ \ \ \ &
\nonumber \\
& \text{\#   NT: \# of t-steps for $Q^2$ dependence}  & \ \ \ \ \ \ &
\nonumber \\
& \text{\#   NX: \# of x-steps for $xD(x)$}           & \ \ \ \ \ \ &
\nonumber \\
& \text{\# NGLI: \# of steps for Gauss-Legendre integral} & \ \ \ \ \ \ &
\nonumber \\
& \text{NT= 580}                            & \ \ \ \ \ \ &
\nonumber \\
& \text{NX= 160}                            & \ \ \ \ \ \ &
\nonumber \\
& \text{NGLI= 32}                           & \ \ \ \ \ \ &
\nonumber \\
& \text{\# NF at the initial scale Q02}     & \ \ \ \ \ \ &
\nonumber \\
& \text{NF= 3}                              & \ \ \ \ \ \ &
\label{eqn:fort10}
\end{alignat}

The parameter IORDER indicates the LO or NLO of $\alpha_s$.
Both LO and NLO evolutions are possible so that the order
of $\alpha_s$ should be IORDER=1 or 2. The scale parameter $\Lambda$
should be supplied in the case of four flavors ($\Lambda_4$).
It is then converted to the three and five flavor values
($\Lambda_3$ and $\Lambda_5$) within the evolution code 
\cite{roberts-book}.
The charm and bottom functions appear as finite distributions
above the threshold values $Q^2>m_c^2$ (or $m_b^2$).
The HQTHRE values are these heavy-quark thresholds
\cite{threshold}.

The kinematical regions of $x$ and $Q^2$ are specified by 
the parameters, $Q_0^2$ (=$Q_{min}^2$), $Q_{max}^2$, and $x_{min}$.
The $Q_0^2$ is the initial $Q^2$ scale,
and $Q_{max}^2$ is the maximum $Q^2$ for calculating the FFs.
Any $Q^2$ values can be chosen as long as perturbative QCD
calculations are valid, which means that small $Q_0^2$ and
$Q_{max}^2$ values are not favorable particularly 
in the region $Q^2<1$ GeV$^2$ where pQCD calculations
do not converge easily due to the large running coupling
constant $\alpha_s$. 
The $Q^2$ maximum $Q^2_{max}=1.558 \times 10^8$ GeV$^2$
is used in making the HKNS07 library for their FFs \cite{hkns07}.
It is chosen so that two grid points are close to the
charm and bottom threshold values $m_c^2$ and $m_b^2$.
If one needs to calculate the evolution only to 
$Q^2$=100 GeV$^2$, one does not have to take such
a large $Q^2_{max}$, and $Q^2_{max}$=100 GeV$^2$ is enough.
We also should note that a very small value of $x_{min}$
is not favored because some FFs become negative, 
which is not physically allowed in principle.
This occurs due to a singular behavior of a time-like
splitting function. In order to cure this issue, more
detailed studies are needed by including resummation effects
\cite{resum}.

The parameter NT is the number of points of $\ln Q^2$ 
in the range $\ln Q^2_{min} \le \ln Q^2 \le \ln Q^2_{max}$,
and NX is the number of points of $x$ in the range 
$x_{min} \le x \le 1$. The NGLI is the Gauss-Legendre points 
for numerical integration.
In the example of Eq. (\ref{eqn:fort10}), the number of 
$t=\ln Q^2$ points is 580, the one of $x$ is 160, and
the one of the Gauss-Legendre points is 32. They are selected
by looking at evolution results by varying their values.
Such studies are explained in details in Sec. \ref{results}.
The NF is the number of flavors at $Q_0^2$, and it is usually
three as given in Eq. (\ref{eqn:fort10}).

\subsection{Sample code}
\label{test}

A test code (sample.f) is provided as an example for running
the main evolution code. The main evolution routine 
GETFF(Q2,X,FF) is called by supplying $Q^2$ and $x$ values.
Evolved FFs are returned to FF(I), (I=$-$5, 5):
\begin{alignat}{6}
& FF(-5)= D_{\bar b}^h (x,Q^2), & \ \ \ 
& FF(-4)= D_{\bar c}^h (x,Q^2), & \ \ \ 
& FF(-3)= D_{\bar s}^h (x,Q^2), & \ \ \ 
& FF(-2)= D_{\bar u}^h (x,Q^2),
\nonumber \\
& FF(-1)= D_{\bar d}^h (x,Q^2), & \ \ \ 
& FF(0)= D_g^h (x,Q^2),         & \ \ \ 
& FF(1)= D_d^h (x,Q^2),         & \ \ \ 
& FF(2)= D_u^h (x,Q^2),         & 
\nonumber \\
& FF(3)= D_s^h (x,Q^2),         & \ \ \ 
& FF(4)= D_c^h (x,Q^2),         & \ \ \ 
& FF(5)= D_b^h (x,Q^2).         & 
& \ \                           & \ \ \ 
\end{alignat}

\section{Results}
\label{results}

\begin{figure}[h!]
   \begin{center}
       \epsfig{file=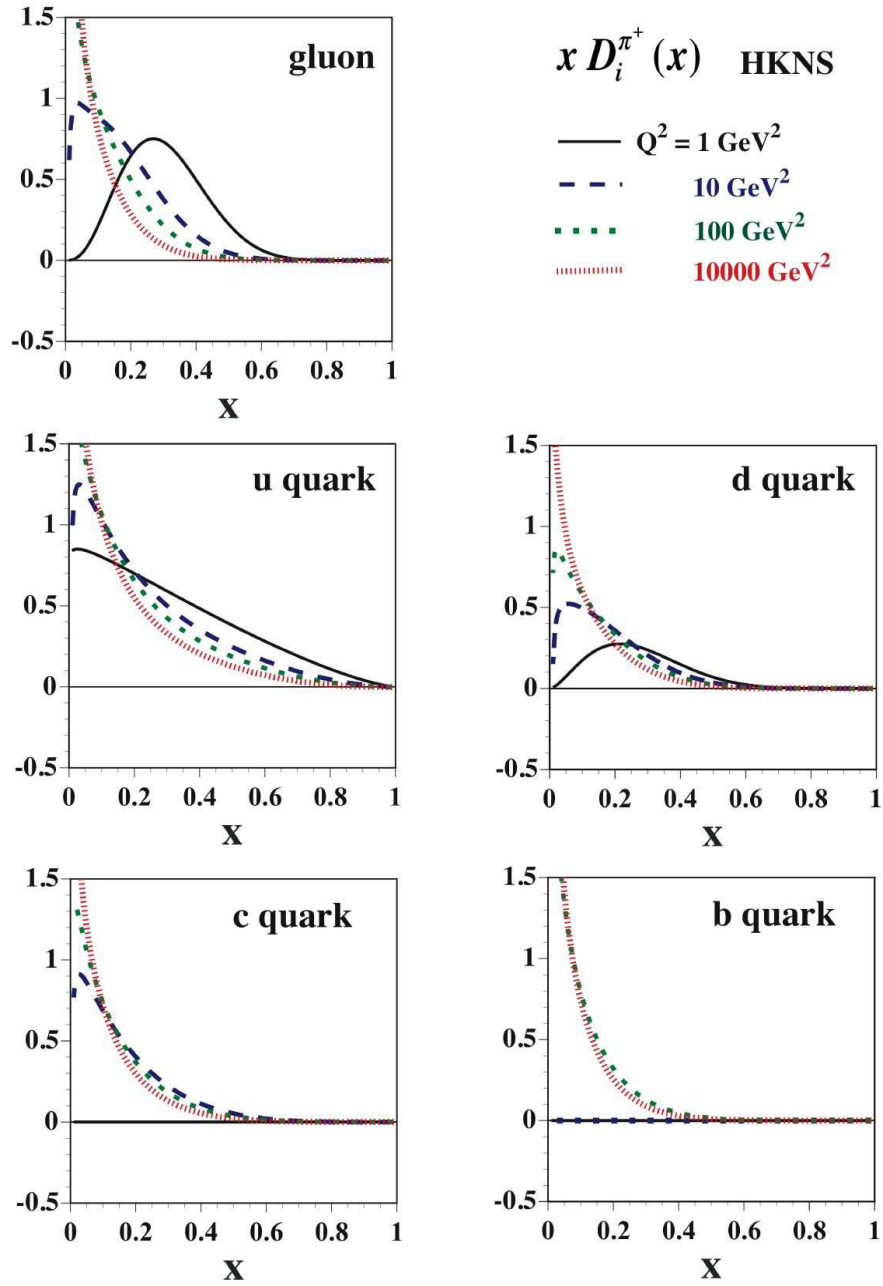,width=0.88\textwidth} 
   \end{center}
\vspace{-0.4cm}
\caption{$Q^2$ evolution of HKNS07 fragmentation 
     functions. The initial $g$, $u$, and $d$
     FFs are supplied at the scale $Q_0^2$=1 GeV$^2$, and
     the $c$ and $b$ functions are at $m_c^2$ and $b_b^2$.
     They are evolved to the scale $Q^2$=10, 100, and 10000 GeV$^2$
     by the time-like DGLAP evolution equations in the NLO
     ($\overline{MS}$) by using the code developed 
     in this work. The explicit parameter values are listed
     in Eq. (\ref{eqn:fort10}).}
\label{fig:hkns07-evol}
\end{figure}

\begin{figure}[h!]
   \begin{center}
       \epsfig{file=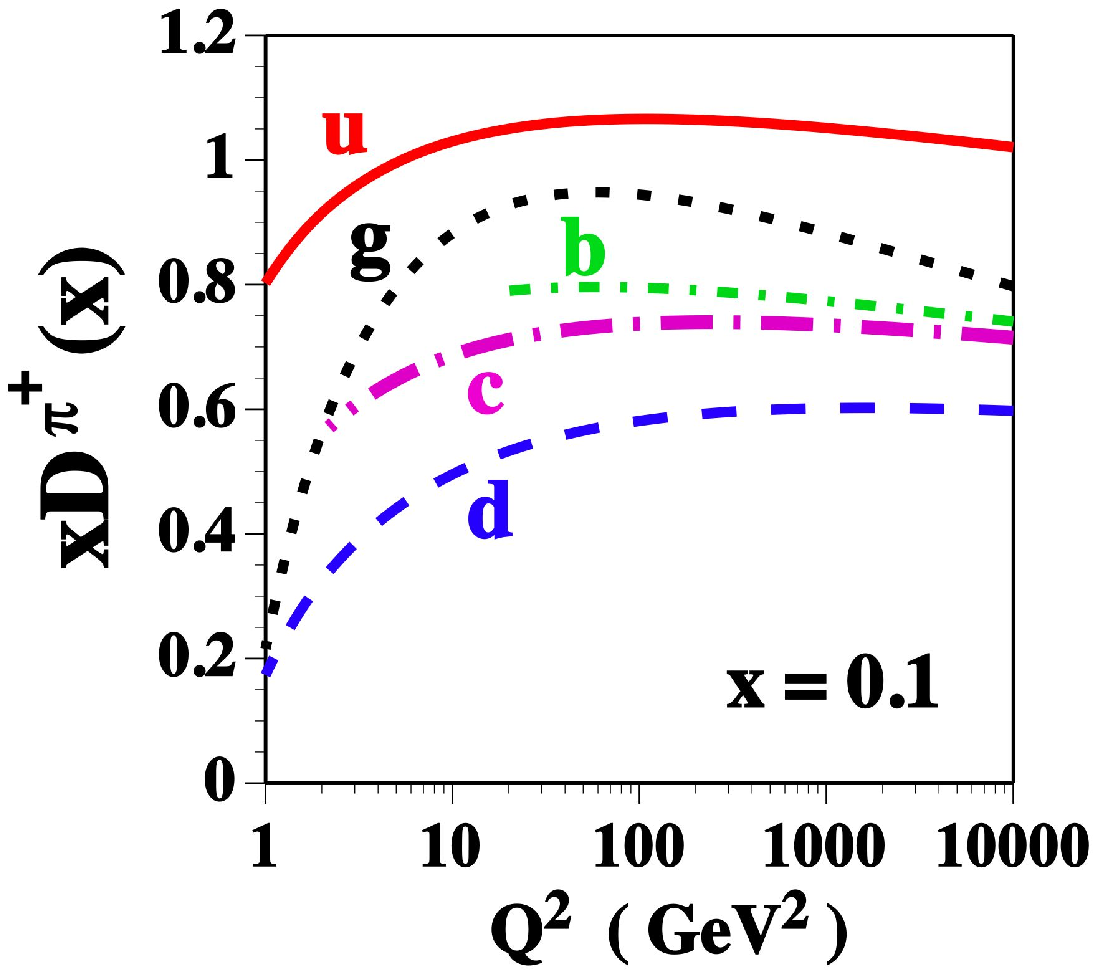,width=0.45\textwidth} 
       \vspace{0.3cm}
       \epsfig{file=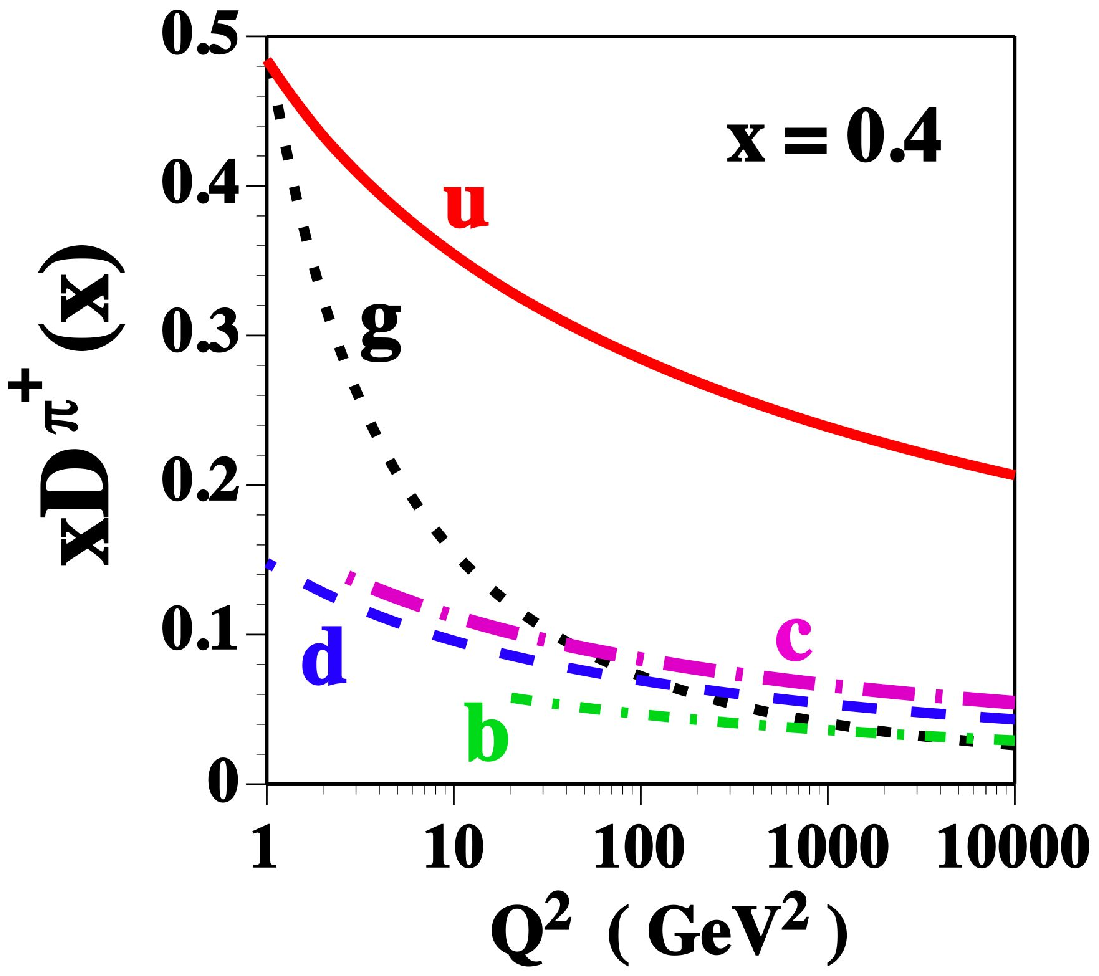,width=0.45\textwidth} 
   \end{center}
\vspace{-0.6cm}
\caption{$Q^2$ dependence of the fragmentation functions
     at fiexed $x$ ($=0.1$ and 0.4) .
     The initial $g$, $u$, and $d$
     FFs are supplied at the scale $Q_0^2$=1 GeV$^2$, and
     the $c$ and $b$ functions are at $m_c^2$ and $b_b^2$.
     They are evolved to the scale $Q^2$=10000 GeV$^2$
     by the time-like DGLAP evolution equations in the NLO
     ($\overline{MS}$) by using the code developed 
     in this work. The explicit parameter values are listed
     in Eq. (\ref{eqn:fort10}).}
\label{fig:hkns07-evol-q2}
\end{figure}

$Q^2$ evolution results of FFs are shown in Fig. \ref{fig:hkns07-evol}
by taking the initial functions of $\pi^+$ as the HKNS07
(Hirai, Kumano, Nagai, Sudoh) parametrization in 2007 \cite{hkns07}. 
The initial functions are provided at $Q^2$=1 GeV$^2$
for $g$, $u$, $d$, and $s$ FFs and for $c$ and $b$ 
at $Q^2=m_c^2$ and $m_b^2$, respectively. The evolution has been
calculated in the NLO and with the scale parameter 
$\Lambda_{QCD}$=0.323 GeV in the running coupling constant.
The used numbers of steps are $N_t$=560, $N_x$=160, and $N_{GL}$=32
for calculating the evolutions.
In Fig. \ref{fig:hkns07-evol-q2}, the $Q^2$ evolution results
are shown as a function of $Q^2$ at fixed $x$ points
($x=0.1$ and 0.4). The same input parameters are used
in setup.ini, which was used in obtaining the results
in Fig. \ref{fig:hkns07-evol}, for running the code.

The input file setup.ini for calculating the evolution
of the NLO HKNS07 functions is given in Eq. (\ref{eqn:fort10}).
The light-parton 
($g$, $u$, $d$, $s$, $\bar u$, $\bar d$, $\bar s$)
FFs are supplied at the initial scale $Q_0^2$, which
is assigned to be the $Q^2$ minimum $Q^2_{min}$.
The evolved $Q^2$ value (10, 100, or 10000 GeV$^2$) needs
to be supplied when running the code, for example, sample.f.

Next, evolution results are shown by varying the parameters
$N_{GL}$, $N_x$, and $N_t$, which affect the evolution 
accuracy. First, the $Q^2$ evolution results are calculated
at $Q^2$=100 GeV$^2$ by using the HKNS07 parametrization
for the initial functions and the parameters
$N_{GL}$=100, $N_x$=500, and $N_t$=500. 
Then, they are considered to be ``standard" functions in
showing ratios with other evolution results.
In the input setup.ini file, $Q^2_{max}$=100 GeV$^2$ is taken
because the larger-$Q^2$ region is not necessary.

\begin{figure}[h!]
   \begin{center}
       \epsfig{file=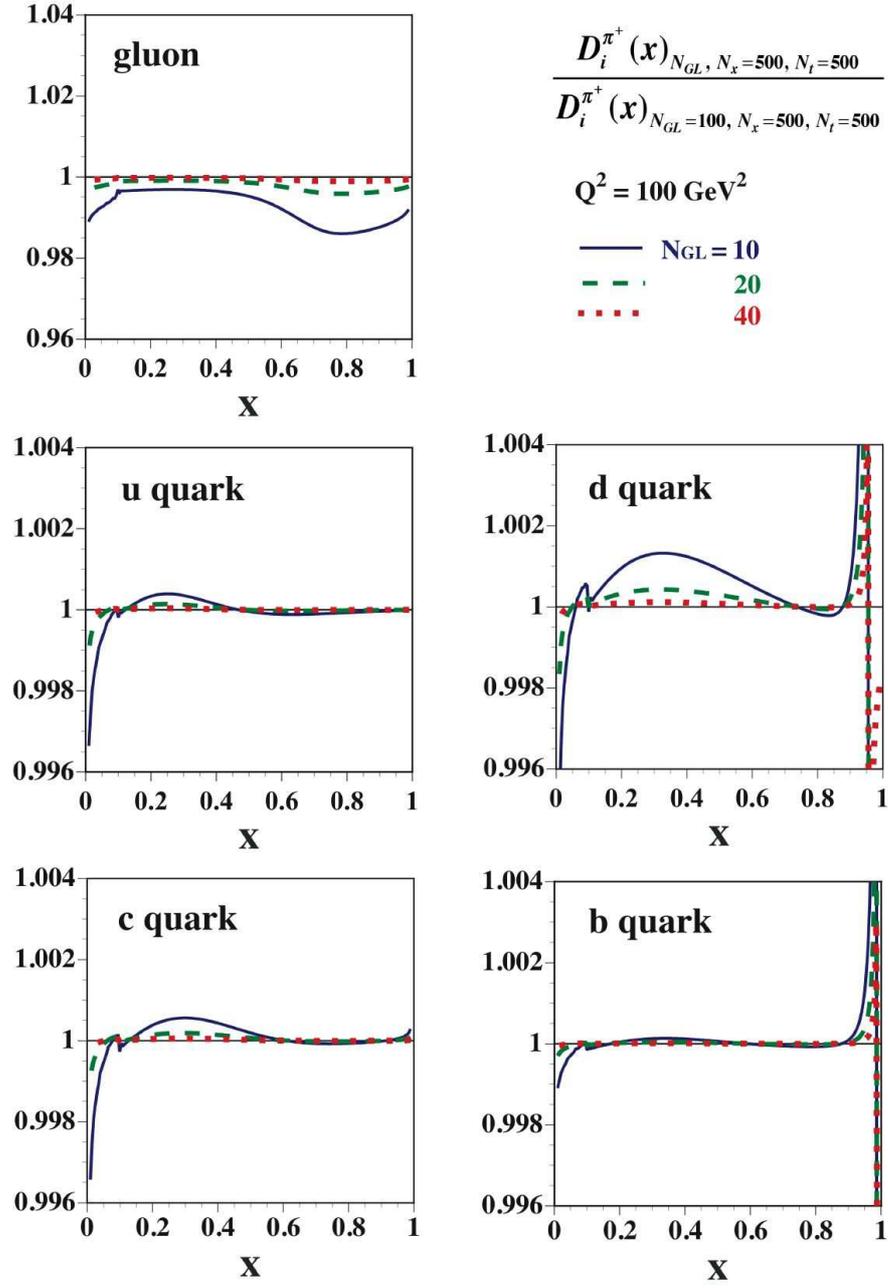,width=0.95\textwidth} 
   \end{center}
\vspace{-0.4cm}
       \caption{\footnotesize
                Evolved fragmentation-function ratios 
                $D_i ^{\pi^+}(x)_{N_{GL},\ N_x=500,\  N_t=500}
                /D_i ^{\pi^+}(x)_{N_{GL}=100,\ N_x=500,\  N_t=500}$
                are shown for $N_{GL}$=10, 20, and 50 
                at $Q^2$=100 GeV$^2$. The initial functions are the HKNS07
                ones at $Q^2$=1 GeV$^2$ for $g$, $u$, and $d$,
                at $Q^2=m_c^2$ for $c$, and at $Q^2=m_b^2$ for $b$.}
       \label{fig:ngl}
\end{figure}

First, $N_{GL}$ is varied as 10, 20, and 40 in order to
find its dependence on evolution results in checking
evolution accuracy. The evolved functions are then used
for calculating ratios with the standard evolution by
$  D_i^{\pi^+} (x,Q^2=100 \text{ GeV}^2)_{N_{GL},N_x=500,N_t=500} 
 / D_i^{\pi^+} (x,Q^2=100 \text{ GeV}^2)_{N_{GL}=100,N_x=500,N_t=500}$.
The ratios are shown in Fig. \ref{fig:ngl} for the fragmentation
functions of $g$, $u$, $d$, $c$, and $b$. The quark functions are
evolved accurately except for the region close to $x=1$
even with a small number of Gauss-Legendre points such as $N_{GL}=10$.
However, the gluon evolution depends much on the choice of $N_{GL}$,
and the results indicate that $N_{GL} \ge 20$ needs to be taken 
for getting the evolution accuracy better than about 0.3\%.
This is the reason why $NGLI=32$ is used in calculating 
the evolutions in Fig. \ref{fig:hkns07-evol}.

\begin{figure}[h!]
   \begin{center}
   \epsfig{file=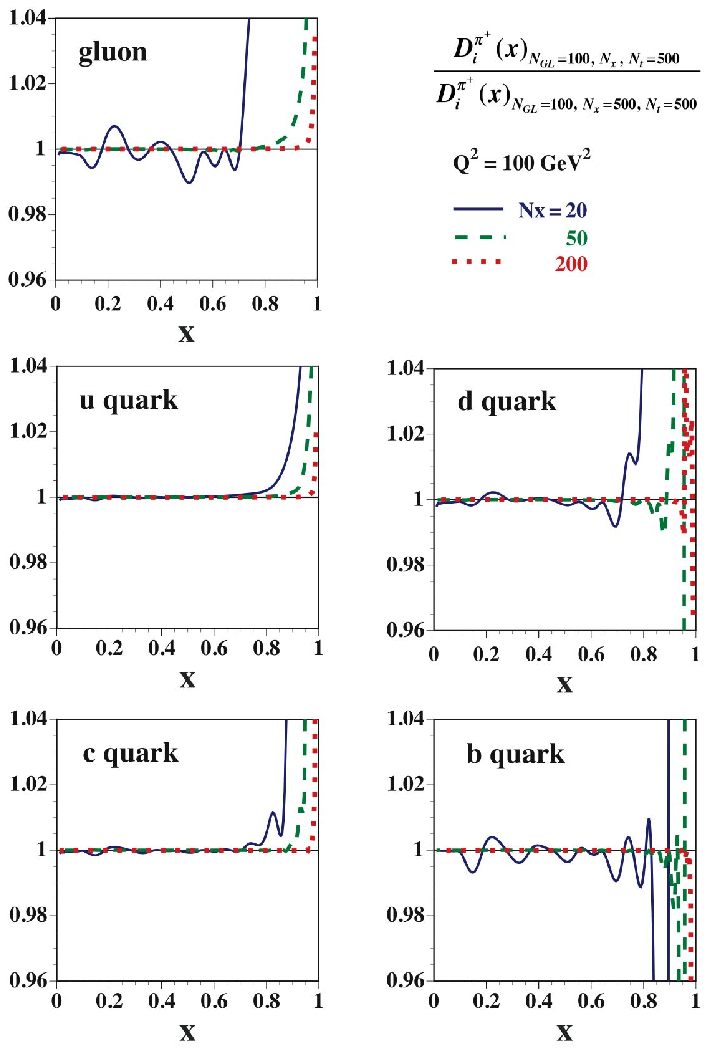,width=0.95\textwidth}
   \end{center}
   \vspace{-0.4cm}
       \caption{\footnotesize
                Evolved fragmentation-function ratios 
                $D_i ^{\pi^+}(x)_{N_{GL}=100,\ N_x,\  N_t=500}
                /D_i ^{\pi^+}(x)_{N_{GL}=100,\ N_x=500,\  N_t=500}$
                are shown for $N_{x}$=20, 50, and 200 
                at $Q^2$=100 GeV$^2$. The other conditions
                are the same as the ones in Fig. \ref{fig:ngl}.}
       \label{fig:nx}
\vspace{-0.0cm}
\end{figure}

Second, the dependence on $N_x$ is shown by fixing the other parameters
at $N_{GL}$=100 and $N_t$=500. The evolved functions are used for 
taking ratios with the standard evolution results with
$N_{GL}$=100, $N_x$=500, and $N_t$=500 in the same way
with Fig. \ref{fig:ngl}. The input parameter $N_x$ is
the number of points in $x$ from $x_{min}$ to one. 
For example, if $x_{min}$=0.01 and $N_x$=500 are taken,
250 points are given for the logarithmic $x$ in the region 
$0.01 \le x \le 0.1$ and another 250 points for the linear $x$ 
in $0.1 \le x \le 1$.
If $x_{min}$=0.001 and $N_x$=600 are taken, we have 400 (200) points
in $0.001 \le x \le 0.1$ ($0.1 \le x \le 1$).
It is changed as $N_x$=20, 50, and 200 to show variations
in the evolved functions, and results are shown in 
Fig. \ref{fig:nx}. In general, there are large differences
in the large-$x$ region in all the FFs. In particular,
the evolved functions are not reliable at $x>0.7$ if 
$N_x$=20 is taken. As the number increases as $N_x$=50 and 200,
they become reliable except for the extremely large-$x$ region 
($x>0.9$). From these studies, $N_x$=160 is taken, for example, 
in Fig. \ref{fig:hkns07-evol} as a reasonable choice.

\begin{figure}[h!]
   \begin{center}
       \epsfig{file=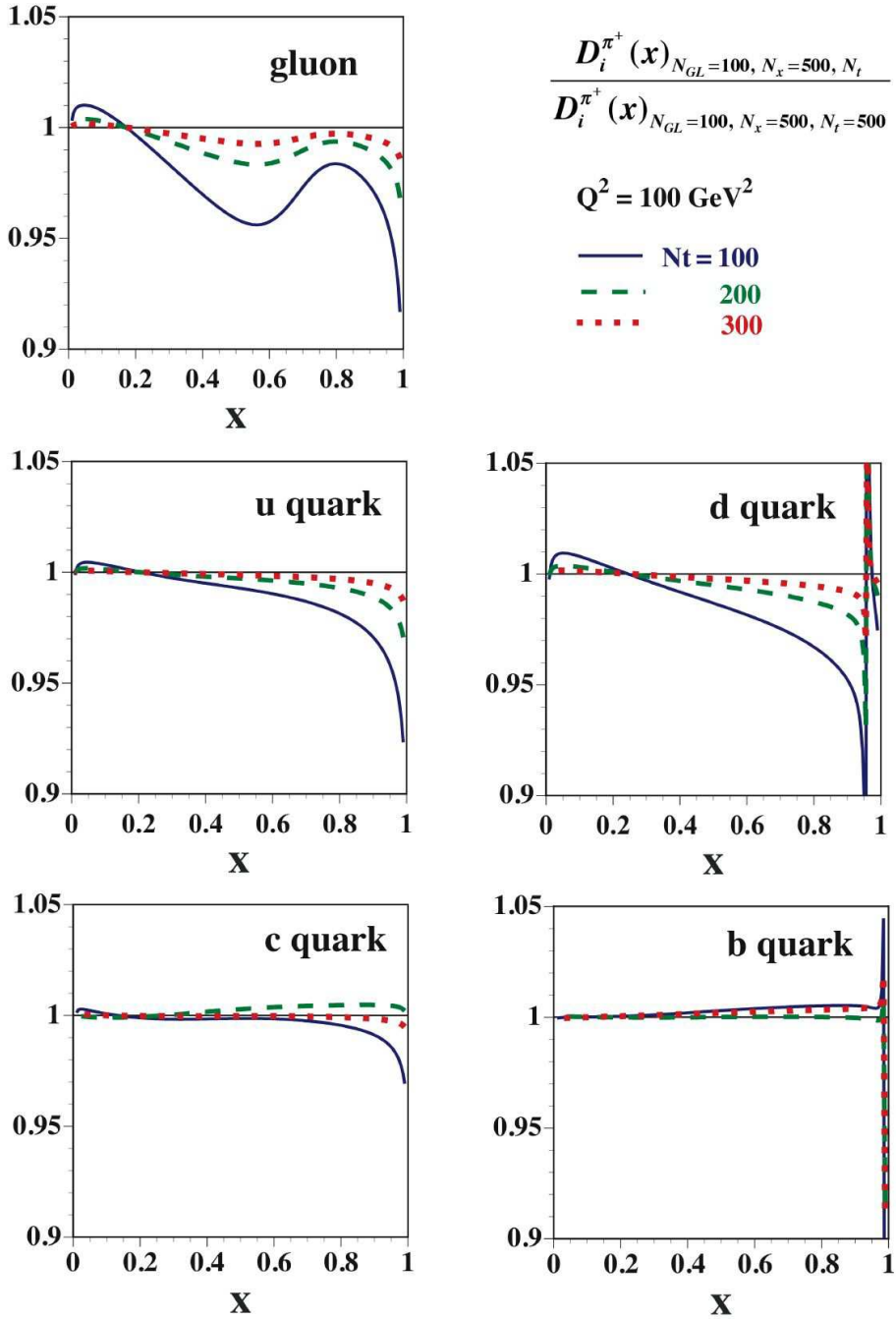,width=0.92\textwidth} 
   \end{center}
   \vspace{-0.4cm}
       \caption{\footnotesize
                Evolved fragmentation-function ratios 
                $D_i ^{\pi^+}(x)_{N_{GL}=100,\ N_x=500,\  N_t}
                /D_i ^{\pi^+}(x)_{N_{GL}=100,\ N_x=500,\  N_t=500}$
                are shown for $N_{x}$=100, 200, and 300 
                at $Q^2$=100 GeV$^2$. The other conditions
                are the same as the ones in Fig. \ref{fig:ngl}.}
       \label{fig:nt}
\vspace{-0.0cm}
\end{figure}

Third, we show $N_t$ dependence in Fig. \ref{fig:nt}.
It is varied as $N_t$=100, 200, and 300 by fixing other
parameters at $N_{GL}=100$ and $N_x=500$.
If $N_t$ is small, evolved distributions are not accurate
at large $x$, especially in the gluon fragmentation function.
A large number of points should be taken for $N_t$ for getting
a converging function within a few percent level of accuracy,
and $N_t$=580 is taken in Fig. \ref{fig:hkns07-evol}.
However, if $Q^2_{max}$ is small, accurate evolution results
can be obtained by taking smaller $N_t$.

A typical running time for obtaining the evolutions
in Fig. \ref{fig:hkns07-evol} 
is 4 seconds by using g95 on the CPU 
(Dual-Core Intel Xeon 2.66 GHz) with Mac-OSX-10.5.8,
so that the code is efficient enough to be used 
on any machines for one's studies on the fragmentation
functions.

\vfill\eject

\section{Summary}
\label{summary}

The fragmentation functions are used in describing hadron-production
cross sections at high energies. The FFs are described by two
variables $x$ and $Q^2$. The $Q^2$ dependence of the FFs
is calculated in perturbative QCD and they are described by
the DGLAP evolution equations. In this work, the $Q^2$ evolution
equations are numerically solved and a useful evolution code
is provided so that other researchers could use it for their 
own studies.
The variables $x$ and $\ln Q^2$ are divided into small steps,
and the evolution is numerically calculated by using 
the Euler method and the Gauss-Legendre quadrature. We showed
that the evolution is accurately calculated except for 
the extremely large-$x$ region by taking reasonably large
numbers of the Gauss-Legendre points ($N_{GL}$), 
$x$ steps ($N_x$), and $t=\ln Q^2$ steps ($N_t$).
Our evolution code can be obtained upon request \cite{request}
for using one's studies on the $Q^2$ evolution of the FFs.

\section*{Acknowledgments}

The authors would like to thank communications with 
W. Bentz, I. C. Cloet, and T.-H. Nagai
about $Q^2$ evolution of fragmentation functions.

\section*{Appendix A. Running coupling constant}

The running coupling constants in the leading order (LO)
and next-to-leading order (NLO) are
\begin{align}
\alpha_s^{LO}(Q^2) & = \frac{4\pi}{\beta_0 \ln (Q^2/\Lambda^2)} ,
\\
\alpha_s^{NLO}(Q^2) & = \frac{4\pi}{\beta_0 \ln(Q^2/\Lambda^2)}
\Biggl[1- \frac{\beta_1 \ln \ln(Q^2/\Lambda^2)}
               {\beta_0^2\ln(Q^2/\Lambda^2)}
\Biggr]  ,
\end{align}
where $\Lambda$ is the QCD scale parameter, 
and $\beta_0$ and $\beta_1$ are given by
\begin{equation}
\beta_0={11\over3}C_G-{4 \over 3}T_R N_f , \ \ \
\beta_1={34\over3}C^2_G-{10\over3}C_G N_f-2C_F N_f  ,
\end{equation}
with the color constants
\begin{equation}
C_A=N_c , \ \ \ 
C_F=\frac{N_c^2-1}{2N_c} , \ \ \ 
T_R=\frac{1}{2} .
\end{equation}
In the NLO, $\overline{MS}$ is used for
the renormalization scheme.

\section*{Appendix B. Splitting functions}

The splitting functions are expanded in $\alpha_s$:
\begin{equation} 
 P_{ij}(x,\alpha_s) = P_{ij}^{(0)}(x) 
           + \frac{\alpha_s(Q^2)}{2 \pi} P_{ij}^{(1)}(x) 
           + \cdot\cdot\cdot ,
\end{equation}
where $P_{ij}^{(0)}(x)$ and $P_{ij}^{(1)}(x)$ are LO and NLO 
splitting functions, respectively. Splitting functions in the LO are
the same as the ones for describing the PDF evolution \cite{bfevol}:
\begin{align}
P_{q_i q_j}^{(0)}(x) & = \delta_{ij} C_F\ \left[\ {{1+x^2} \over {(1-x)_+}}
                    \ + \ {3 \over 2} \ \delta (1-x)\ \right] , 
\\
P_{qg}^{(0)}(x) & = T_R \left[\ x^2 \ + \ (1-x)^2\ \right] , 
\\
P_{gq}^{(0)}(x) & = C_F \ {{1+(1-x)^2} \over x}  , 
\\
P_{gg}^{(0)}(x) & = 2C_G \left[\ {x \over {(1-x)_+}} \ + \
                        {{1-x} \over x} \ + \ x(1-x) \ + \
       \left( {{11}\over{12}} - {1 \over 3}{{N_fT_R} \over {C_G}}\right)
       \ \delta (1-x) \ \right] .
\end{align}
The only point one should note is that the splitting functions
$P_{qg}$ and $P_{gq}$ are interchanged in the matrix of 
Eq. (\ref{eqn:evolution-q+}) from the PDF evolution.
However, the spacelike and timelike splitting functions
for the PDFs and FFs, respectively, are different in higher-order
of $\alpha_s$ as shown in Refs. \cite{esw-book,splitting}.
The quark-quark splitting function in the NLO is given by
\begin{align}
P_{q_i^+ q_j^+}^{(1)} & \equiv P_{q_i q_j}^{(1)} + P_{q_i \bar q_j}^{(1)}
                  = \delta_{ij} ( P_{q q}^{V(1)} + P_{q \bar q}^{V(1)} )
                              +   P_{q q}^{S(1)} + P_{q \bar q}^{S(1)} ,
\end{align}
where the functions $P_{q q}^{V(1)}$ and $P_{q \bar q}^{V(1)}$ are
given in Ref. \cite{esw-book}, the function $P_{q q}^{S(1)}$
($=P_{q \bar q}^{S(1)}$) can be derived from the relation 
$P_{qq}^{(1)} = P_{qq}^{V(1)} + P_{q\bar q}^{V(1)}
          +N_f (P_{qq}^{S(1)} + P_{q\bar q}^{S(1)}) $.
These expressions are lengthy and they are provided
in Sec. 6.1 of Ref. \cite{esw-book}.
          


\vfill\eject

\noindent
{\bf sample.f}
\begin{verbatim}
C ---------------------------------------------------------------------
      PROGRAM SAMPLE                                      ! 2011-11-24
C ---------------------------------------------------------------------
C X DEPENDENCE OF FFs 
      IMPLICIT REAL*8(A-H,O-Z)
      PARAMETER(NSTEP=200)
      CHARACTER*1 Q2PROG_END
      CHARACTER*9 Q2_OUT
      CHARACTER*22 FNAME
      DIMENSION FF(-5:5),Z(600)
      COMMON /XMIN/XMIN  ! Setting in the setup.ini

      CALL FF_DGLAP() ! Making arrary of FF(XMIN:1.D0, Q2_ini:Q2_max)
      
200   WRITE(*,fmt='(a)') "Q^2= "; READ(*,*) Q2
      WRITE(Q2_OUT,'(1PE9.3)') Q2
      FNAME='Q2='//Q2_OUT//'_GeV2.dat'
      OPEN(unit=23,file=FNAME,FORM='formatted')

      DLMIN=DLOG10(XMIN)
      ZLSTEP=(DLOG10(1.D0)-DLMIN)/DFLOAT(NSTEP)
      DO I=1,NSTEP+1
        DLOGZ=DFLOAT(I-1)*ZLSTEP+DLMIN
        Z(I)=10.D0**(DLOGZ)
      END DO

C FOR pi^+, FF(I), I= 0:g, 1:d, 2:u, 3:s, 4:c, 5:b
      DO I=1,NSTEP
        CALL GETFF(Q2,Z(I),FF) ! Getting FF(Z,Q^2)
        WRITE(23,1010) Z(I), Z(I)*FF(0), ! gluon
     +                       Z(I)*FF(2), ! up
     +                       Z(I)*FF(1), ! down
     +                       Z(I)*FF(3), ! strange
     +                       Z(I)*FF(4), ! charm
     +                       Z(I)*FF(5)  ! bottom
      END DO
      WRITE(*,fmt='(a)') 
     +      "Do you finish the FF Q2 evolution ? (y/n) "     
      READ(*,*) Q2PROG_END
      IF((Q2PROG_END(1:1).EQ.'n').OR.(Q2PROG_END(1:1).EQ.'N')) GOTO 200

 1010 FORMAT(1X,9(1PE16.7))
      CLOSE(23)
      END
C ---------------------------------------------------------------------
\end{verbatim}

{\bf TEST RUN OUTPUT}

\vspace{0.3cm}

Running the distributed sample code (sample.f) together with
the main $Q^2$ evolution subroutine (FF\_DGLAP.f) and the input
file (setup.ini), we obtain the following output for
$Q^2$=100 GeV$^2$. The following functions corresponds to
the curves at $Q^2$=100 GeV$^2$ in Fig. \ref{fig:hkns07-evol}.

\vspace{0.5cm}
\noindent
\begin{footnotesize}
\begin{tabular}{lllllll} 
\ \ \ \ \ \ \ \ $x$            & \ \ \ \ \ \ $x \, D_g^{\pi^+}$ &
\ \ \ \ \ \ $x \, D_u^{\pi^+}$ & \ \ \ \ \ \ $x \, D_d^{\pi^+}$ & 
\ \ \ \ \ \ $x \, D_s^{\pi^+}$ & \ \ \ \ \ \ $x \, D_c^{\pi^+}$ & 
\ \ \ \ \ \ $x \, D_b^{\pi^+}$ \\ 
\ & \ & \ & \ & \ & \ & \ \\
\hs 1.0000000E-02 &  \hs 1.7387272E+00 &  \hs 1.5206099E+00 &  
\hs 7.1337454E-01 &  \hs 7.1374282E-01 &  \hs 1.2897181E+00 &  
\hs 2.5241504E+00 \\
\hs 1.0232930E-02 &  \hs 1.7469241E+00 &  \hs 1.5297365E+00 &  
\hs 7.2321764E-01 &  \hs 7.2358058E-01 &  \hs 1.2947968E+00 &  
\hs 2.5089220E+00 \\
\hs 1.0471285E-02 &  \hs 1.7544533E+00 &  \hs 1.5384281E+00 &  
\hs 7.3265629E-01 &  \hs 7.3301393E-01 &  \hs 1.2995278E+00 &  
\hs 2.4936096E+00 \\
\hs 1.0715193E-02 &  \hs 1.7613271E+00 &  \hs 1.5466907E+00 &  
\hs 7.4169702E-01 &  \hs 7.4204939E-01 &  \hs 1.3039165E+00 &  
\hs 2.4782151E+00 \\
\hs 1.0964782E-02 &  \hs 1.7675595E+00 &  \hs 1.5545311E+00 &  
\hs 7.5034733E-01 &  \hs 7.5069444E-01 &  \hs 1.3079690E+00 &  
\hs 2.4627404E+00 \\
\hs 1.1220185E-02 &  \hs 1.7731631E+00 &  \hs 1.5619553E+00 &  
\hs 7.5861384E-01 &  \hs 7.5895574E-01 &  \hs 1.3116905E+00 &  
\hs 2.4471874E+00 \\
\hs 1.1481536E-02 &  \hs 1.7781510E+00 &  \hs 1.5689699E+00 &  
\hs 7.6650367E-01 &  \hs 7.6684037E-01 &  \hs 1.3150869E+00 &  
\hs 2.4315579E+00 \\
\hs 1.1748976E-02 &  \hs 1.7825365E+00 &  \hs 1.5755812E+00 &  
\hs 7.7402396E-01 &  \hs 7.7435550E-01 &  \hs 1.3181639E+00 &  
\hs 2.4158538E+00 \\
\hs 1.2022644E-02 &  \hs 1.7863321E+00 &  \hs 1.5817955E+00 &  
\hs 7.8118145E-01 &  \hs 7.8150786E-01 &  \hs 1.3209270E+00 &  
\hs 2.4000771E+00 \\
\hs 1.2302688E-02 &  \hs 1.7895498E+00 &  \hs 1.5876182E+00 &  
\hs 7.8798236E-01 &  \hs 7.8830368E-01 &  \hs 1.3233809E+00 &  
\hs 2.3842291E+00 \\
\hss $\cdot\cdot\cdot$ &  \hss $\cdot\cdot\cdot$ & 
\hss $\cdot\cdot\cdot$ &  \hss $\cdot\cdot\cdot$ & 
\hss $\cdot\cdot\cdot$ &  \hss $\cdot\cdot\cdot$ & 
\hss $\cdot\cdot\cdot$ \\
\hss $\cdot\cdot\cdot$ &  \hss $\cdot\cdot\cdot$ & 
\hss $\cdot\cdot\cdot$ &  \hss $\cdot\cdot\cdot$ & 
\hss $\cdot\cdot\cdot$ &  \hss $\cdot\cdot\cdot$ & 
\hss $\cdot\cdot\cdot$ \\
\hss $\cdot\cdot\cdot$ &  \hss $\cdot\cdot\cdot$ & 
\hss $\cdot\cdot\cdot$ &  \hss $\cdot\cdot\cdot$ & 
\hss $\cdot\cdot\cdot$ &  \hss $\cdot\cdot\cdot$ & 
\hss $\cdot\cdot\cdot$ \\
\hs 8.9125094E-01 &  \hs 2.7586761E-05 &  \hs 7.8509439E-03 &  
\hs 1.0318809E-06 &  \hs 1.0318809E-06 &  \hs 4.2783416E-05 &  
\hs 1.7161909E-06 \\
\hs 9.1201084E-01 &  \hs 1.4513488E-05 &  \hs 5.0877079E-03 &  
\hs 2.3777347E-07 &  \hs 2.3777347E-07 &  \hs 1.6972984E-05 &  
\hs 4.9828819E-07 \\
\hs 9.3325430E-01 &  \hs 6.3142094E-06 &  \hs 2.8910396E-03 &  
\hs 3.2396330E-08 &  \hs 3.2396330E-08 &  \hs 5.0850503E-06 &  
\hs 9.9610221E-08 \\
\hs 9.5499259E-01 &  \hs 1.9415364E-06 &  \hs 1.2925565E-03 &  
\hs 7.2014456E-10 &  \hs 7.2014458E-10 &  \hs 9.1254976E-07 &  
\hs 1.0039050E-08 \\
\hs 9.7723722E-01 &  \hs 2.5705072E-07 &  \hs 3.2218983E-04 &  
\hs$-$2.2238409E-10 &  \hs $-$2.2238409E-10 &  \hs  4.7361837E-08 &  
\hs  1.7555729E-10 \\
\end{tabular}
\end{footnotesize}

\end{document}